\documentclass[aps,prl,twocolumn,groupedaddress,showpacs]{revtex4}
\usepackage{graphicx}
\usepackage{dcolumn}
\usepackage{bm}
\usepackage{epsfig}
\input epsf
\epsfclipon

\begin{document}
\title{Universal scaling of distances in complex networks}
\author{Janusz A. Ho{\l}yst, Julian Sienkiewicz, Agata Fronczak, Piotr Fronczak, and Krzysztof
Suchecki} \affiliation{Faculty of Physics and Center of Excellence
for Complex Systems Research, Warsaw University of Technology,
Koszykowa 75, PL-00-662 Warsaw, Poland}
\date{\today}
\begin{abstract}
Universal scaling of distances between vertices of
Erd\H{o}s-R\'{e}nyi random graphs, scale-free Barab\'asi-Albert
models, science collaboration networks,  biological networks,
Internet Autonomous Systems and public transport networks are
observed. A mean distance between two  nodes of degrees $k_i$ and
$k_j$ equals to $\langle l_{ij}\rangle=A-B\log(k_i k_j)$. The
scaling is valid over several decades. A simple theory for the
appearance of this scaling is presented. Parameters $A$ and $B$
depend on the mean value of a node degree  $\langle k \rangle
_{nn}$ calculated for the nearest neighbors and on network clustering
coefficients.
\end{abstract}  \maketitle

\par Recently, much effort has been put into
investigation of network systems, in order to recognize their
structures and emerging complex properties (for a review see
\cite{BArev,0a,0b,Pastor_book}). The empirical analysis of many real complex
networks has revealed the presence of several universal scaling
laws. The scale-free behavior of degree distributions $P(k)\sim
k^{-\gamma}$ \cite{BAscience} observed in a number of social,
biological and technological systems is probably the most amazing.
Aside from that, many further scaling laws have been found, such as
a dependence of clustering coefficient on node degree in
hierarchical networks $c(k)\sim k^{-\alpha}$ \cite{ck}, scaling of
connection weight distribution \cite{str1,str2}, connection load
distribution \cite{load}, load dependence on degree
\cite{betweenness} and others \cite{community,rivers,allom}.

At the macro-scale one can describe a whole network by a
dependence of a mean distance between any pair of nodes on the
network size and in many real networks the small-world effect is observed \cite{watts},
i.e. the mean distance $l$ between nodes of such
networks increases not faster than logarithmically with their
size $N$. In scale-free networks the small-world effect changes
to ultra small-world effect ($l\sim \log\log N$) when $\lambda<3$
\cite{havlin,metric,Fronczak1,Fronczak2}. It was also observed
that if a network disorder is present, optimal paths become much
longer and the small-world effect disappears \cite{Braunstein}.
The recent research on complex networks is slowly shifting from
problems of network topology to directed and weighted networks
\cite{PRLbarrat,Krap,Dor}, network dynamics \cite{PRLBarab}, as
well as to the issue of network efficiency \cite{Newman04}.

In the present paper we come back to networks geometry and analyze
surprising empirical scaling that has not been considered before. We
think that our observations can be important for understanding of
network structures and for processes driving their evolution as
well as for constructing search algorithms in real web-like systems.
We show that the mean distance between nodes with
degrees $k_{i}$ and $k_{j}$ is given by the following relation
\begin{equation}\label{law0}
\langle l_{ij}\rangle=A-B\log (k_{i} k_{j}).
\end{equation}
The above scaling law is shown to be correct not only for network models
 but also for many real networks regardless of their degree distribution and correlation
profiles.

\begin{figure*}
 \centerline{\epsfig{file=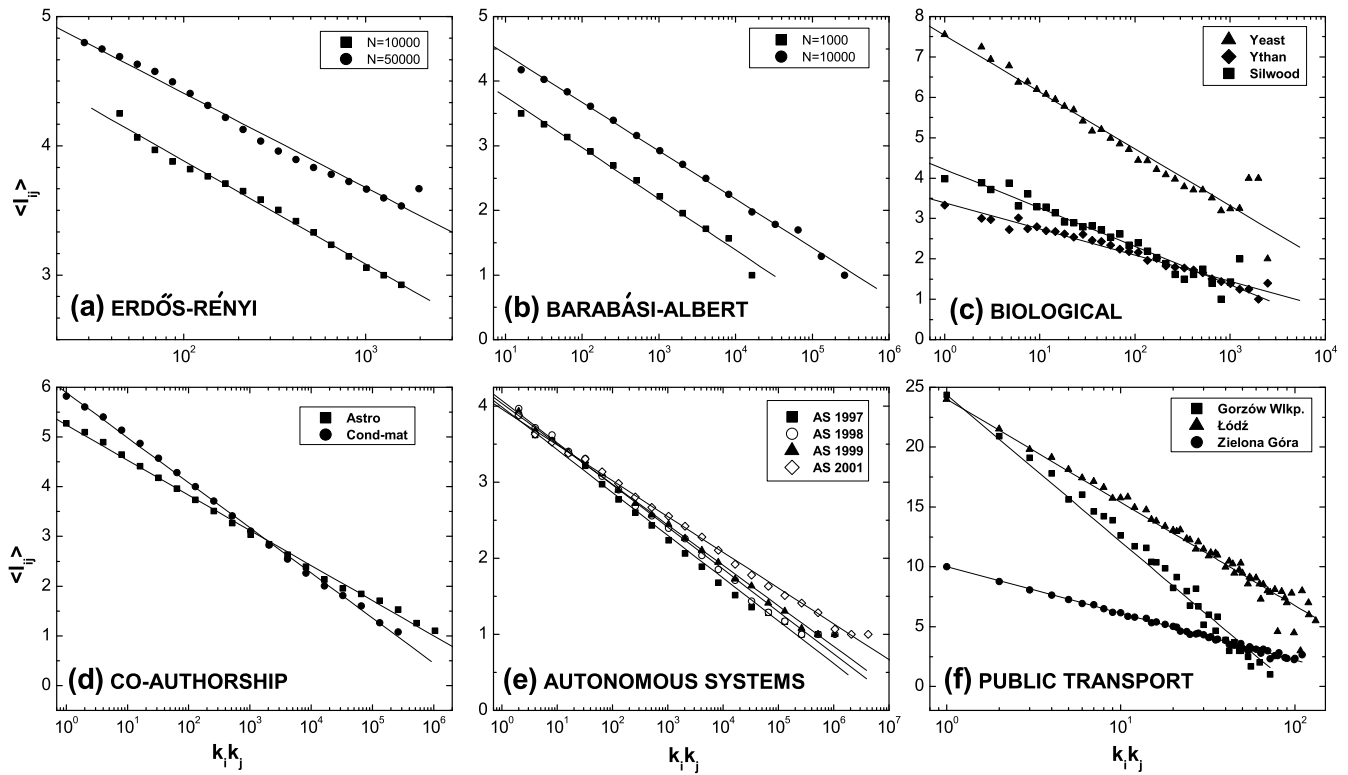,angle=0,width=\textwidth}}
    \caption{Mean distance $\langle l_{ij}\rangle$
between pairs of nodes $i$ and $j$ as a function of a product of
their degrees $k_i k_j$.
    \textbf{(a)} \emph{Erd\H{o}s-R\'enyi} random graphs: $\langle k \rangle=20$ and $N=10000$ (squares)
     $N=50000$ (circles),
     \textbf{(b)}\emph{Barab\'asi-Albert} networks: $\langle k \rangle=8$ and $N=1000$ (squares) $N=10000$
    (circles),
     \textbf{(c)} Biological networks: \emph{Silwood} (squares), \emph{Yeast} (triangles), \emph{Ythan}
     (diamonds),
     \textbf{(d)} Co-authorship networks: \emph{Astro} (squares), \emph{Cond-mat}
     (circles),
     \textbf{(e)} Internet Autonomous Systems: \emph{Year 1997} (squares), \emph{Year 1998} (circles)
    \emph{Year 1999} (triangles), \emph{Year 2001} (diamonds),
    \textbf{(f)} Public transport networks in Polish cities: \emph{Gorz\'ow Wlkp.} (squares),
    \emph{{\L}\'od\'z} (triangles), \emph{Zielona G\'ora} (circles)
     In \textbf{(b)}, \textbf{(d}) and \textbf{(e)} data are logarithmically binned with the power of 2,
     in case of \textbf{(c)} and \textbf{(a)} with the power of 1.25 and in case of \textbf{(f)} data are not binned.}
    \label{fig_all}
\end{figure*}

Fig. \ref{fig_all} presents mean distance $\langle
l_{ij}\rangle$ between pairs of nodes $i$ and $j$ as a function of
a product of their degrees $k_i k_j$ in selected complex networks.
Analyzed systems belong to
very different classes ranging from generic models of random
graphs and scale-free networks, through natural systems such as
food webs and metabolic networks to man-made like the Internet and
public transport networks. We include data for Erd\H{o}s-R\'enyi
random graphs, Barab\'asi-Albert evolving networks, biological
networks \cite{infoBiol,food,yeast} ({\it Silwood}, {\it Ythan}, {\it Yeast}), social
networks \cite{infoCol,newman01a} (co-authorship groups {\it Astro} and {\it Cond-mat}),
Internet Autonomous Systems \cite{infoInt}
and selected networks for public transport in Polish cities \cite{infoTran,Julek}
(Gorz\'ow Wlkp., {\L}\'od\'z, Zielona G\'ora). One can see,
that the relation (\ref{law0}) is very well fulfilled over
several decades for all our data. Let us stress that the networks
mentioned above display a wide variety of basic characteristics.
Among them there are both scale-free and single scale networks,
with either negligible or very high clustering coefficient,
assortative \cite{correlations}, disassortative or uncorrelated.
The only apparent common feature of all above systems is the
small-world effect. We have checked however that for the
small-world Watts-Strogatz model \cite{watts}, the scaling (\ref{law0})
is nearly absent and it is visible only for large rewiring probability,
and only for larger degrees, where nodes have many shortcuts.

\begin{figure}
\vskip .5cm
\centerline{\epsfig{file=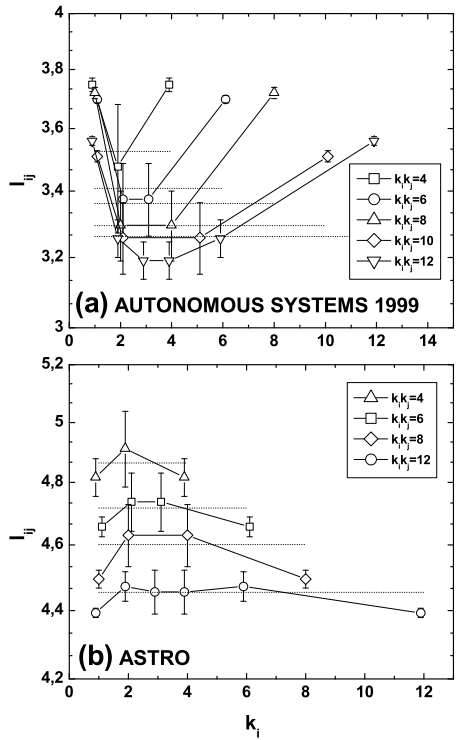,width=.71\columnwidth,angle=0}}
\caption{\label{graph_KS} Dependence of average path length on
$k_i$, for fixed $k_i k_j$ product. The lines
connecting the symbols are there for clarity. The bars show point
weight, meaning relative numbers of pairs $ij$. The horizontal
lines are weighted averages over $k_i$, and are just average path
lengths for given $k_i k_j$. The top shows data for the Internet
Autonomous Systems, while bottom for {\it Astro} co-authorship network.
Note: The very small shifts on $k_i$ axis between data for
different $k_i k_j$ are artificially introduced to make the weight
bars not overlap.}
\end{figure}

Although the scaling (\ref{law0}) works well for distances
averaged over all pairs of nodes specified by a given product $k_i
k_j$ there can be large differences if one changes $k_i$ while
keeping $k_i k_j$ constant. The Fig. \ref{graph_KS} presents the
dependence of average path length $l_{ij}$ on $k_i$, for a fixed
product $k_i k_j$ in the case of {\it Astro} network and for the
Internet Autonomous Systems in 1999. One can see that although the
{\it Astro} network is assortative \cite{correlations} (short-range attraction), pairs of
nodes with similar degrees are in average further away than
different degree pairs (long-range repulsion). For the
disassortative network AS \cite{correlations} the behavior is opposite. For
uncorrelated networks (Erd\H{o}s-R\'enyi, Barab\'asi-Albert), the
average path length is constant if the product $k_i k_j$ is fixed
\cite{nasz_PhysicaA}.

To justify the relation (\ref{law0})  let us consider a
simple approach that bases on a concept of branching trees
exploring the space of a random network. We need to estimate the
mean shortest path between a node $i$ of degree $k_i$ and a node
$j$ of degree $k_j$. Let us notice that
following a random direction of a randomly chosen edge one
approaches node $j$ with a probability $p_j=k_j/(2E)$, where $2E=N
\langle k \rangle$ is a double number of links. It follows that in
average one needs
$M_{j}=1/p_{j}=2E/k_j$
of random trials to arrive at the node $j$.

\begin{figure}
 \centerline{\epsfig{file=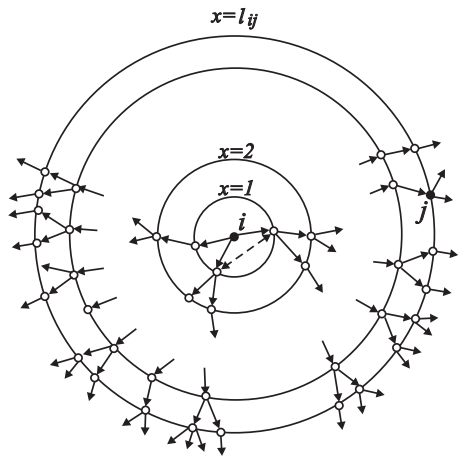,width=.7\columnwidth}}
    \caption{Tree formed by a random process, starting from the node $i$ and approaching the node $j$.}
        \label{fig_1}
\end{figure}

Now let us consider a branching process represented by the tree
$T_i$ (Fig. \ref{fig_1}) that starts at the node $i$ where an
average branching factor is $\kappa$ (all loops are neglected). If
a distance between the node $i$ and the surface of the tree equals
to $x$ then in average there are
$N_i=k_i\kappa^{x-1}$
nodes at such a surface and there is the same number of
links ending at these nodes.  It follows that in average the tree
$T_i$ touches the node $j$ when
$N_i=M_j$ i.e. when
\begin{equation}\label{eq4a}
k_ik_j\kappa^{x-1}=N\langle k\rangle\label{prod}.
\end{equation}
 Since the
mean distance from the node $i$ to the node $j$ is $\langle
l_{ij}\rangle=x$ thus we get the scaling relation (\ref{law0})
with
\begin{equation}\label{e5}
A=1+\frac{\log (N \langle k \rangle)}{\log
\kappa}\;\;\;\;\mbox{and}\;\;\;\; B=\frac{1}{\log \kappa}
\label{AB}.
\end{equation}
The result (\ref{AB}) is in agreement with the paper
\cite{Niemiec} where the concept of generating functions for
random graphs  has been used.

One has to take into account that in the above considerations we
have assumed there are no degree-degree correlations, we have
neglected all loops and we have treated the branching level $x$ as
a continuum variable to fulfill the relation (\ref{eq4a}). The
last approximation can be improved if one finds a probability
distribution for $P(l_{ij})$ and a corresponding average value
$\langle l_{ij}\rangle$. Such an approach  has been performed in
our papers \cite{Fronczak1,Fronczak2} where we have applied the
concept of hidden variables and have received the same value of the
parameter $B$ and small corrections to $A$.

The mean branching factor $\kappa$ is a mean value over all local
branching factors and over all trees in the network. In the first
approximation it could be estimated as the mean arithmetic value
of a nearest neighbor degree less one (incoming edge): $\kappa
=\langle k\rangle_{nn}-1$. Such a mean value is however not exact
because local branching factors in every tree are {\it multiplied}
one by another in (\ref{prod}). The corrected mean value of
$\kappa$ should be taken as an arithmetic mean value over all
geometric values from different trees, what is very difficult to
perform numerically. We calculate arithmetic mean branching factor
over nearest neighborhood of every node $m$, i.e. $\kappa^{(m)}
=\langle k\rangle_{nn}^{(m)}-1$, and then average it geometrically
over all nodes $m$, i.e. $\kappa=\langle \kappa^{(m)}\rangle_m$.
Although our approach is not exact, it leads to a good agreement
between coefficients $A_e$, $B_e$ taken from real networks (see
Table \ref{tab1}) and $A$, $B$ calculated from our model.

\begin{table*}
\setlength{\tabcolsep}{4.5pt}
\begin{center}
\begin{tabular}{lccccccccccccc}
\hline\hline network & $N$ & $\langle k \rangle$ & $c$ & $A_e$ &
$A$ & $\Delta A / A$ & $A'$ & $\Delta A' / A'$ & $B_e$ & $B$ & $\Delta B / B$ & $B'$ & $\Delta B' / B'$\\
\hline
Erd\H{o}s-R\'{e}nyi & 10000 & 20.00 & 0.002 & 5.48 & 5.08 & -0.08 & 5.08 & 0.08 & 0.798 & 0.769 & -0.04 & 0.770 & -0.04\\
Erd\H{o}s-R\'{e}nyi & 50000 & 20.00 & 0.000 & 5.86 & 5.61 & -0.04 & 5.61 & -0.04 & 0.729 & 0.769 & 0.05 & 0.769 & 0.05\\
Barab\'asi-Albert & 1000 & 8.00 & 0.038 & 4.54 & 4.24 & -0.07 & 4.27 & -0.06 & 0.813 & 0.830 & 0.02 & 0.842 & 0.03\\
Barab\'asi-Albert & 10000 & 8.00 & 0.007 & 5.17 & 4.81 & -0.08 & 4.81 & -0.07 & 0.778 & 0.777 & 0.00 & 0.779 & 0.00\\
Astro & 13986 & 25.56 & 0.609 & 5.24 & 4.30 & -0.22 & 4.98 & -0.05 & 0.707 & 0.595 & -0.19 & 0.786 & 0.10\\
Cond-mat & 17013 & 9.46 & 0.604 & 5.90 & 5.09 & -0.16 & 6.38 & 0.08 & 0.908 & 0.786 & -0.16 & 1.150 & 0.21\\
Silwood & 153 & 4.77 & 0.142 & 4.22 & 3.69 & -0.14 & 3.78 & -0.12 & 0.955 & 0.941 & -0.02 & 1.004 & 0.05\\
Yeast & 1846 & 2.39 & 0.068 & 7.53 & 6.66 & -0.13 & 6.87 & -0.10 & 1.406 & 1.552 & 0.09 & 1.629 & 0.14\\
Ythan & 135 & 8.83 & 0.216 & 3.39 & 3.35 & -0.01 & 3.45 & 0.02 & 0.649 & 0.765 & 0.15 & 0.832 & 0.22\\
AS 1997 & 3015 & 3.42 & 0.182 & 3.99 & 3.39 & -0.18 & 3.42 & -0.17 & 0.562 & 0.596 & 0.06 & 0.629 & 0.11\\
AS 1998 & 4180 & 3.72 & 0.250 & 4.08 & 3.41 & -0.20 & 3.45 & -0.18 & 0.555 & 0.575 & 0.04 & 0.620 & 0.10\\
AS 1999 & 5861 & 3.86 & 0.250 & 4.03 & 3.35 & -0.20 & 3.38 & -0.19 & 0.532 & 0.540 & 0.01 & 0.579 & 0.08\\
AS 2001 & 10515 & 4.08 & 0.289 & 3.96 & 3.23 & -0.23 & 3.25 & -0.22 & 0.471 & 0.481 & 0.02 & 0.518 & 0.09\\
Gorz\'ow Wlkp. & 269 & 2.48 & 0.082 & 24.36 & 16.06 & -0.52 & 19.76 & -0.23 & 12.27 & 5.333 & -1.30 & 6.651 & -0.84\\
{\L}\'od\'z & 1023 & 2.83 & 0.065 & 24.01 & 11.67 & -1.06 & 12.70 & -0.89 & 8.621 & 3.084 & -1.80 & 3.389 & -1.54\\
Zielona G\'ora & 312 & 2.98 & 0.067 & 10.03 & 8.96 & -0.12 & 9.63 & -0.04 & 3.908 & 2.682 & -0.46 & 2.917 & -0.34\\
\hline\hline
\end{tabular}
\end{center}
\caption{Comparison between experimental and theoretical data.
\emph{Astro} and \emph{Cond-mat} are co-authorship networks,
 \emph{Silwood},
\emph{Yeast} and \emph{Ythan} are biological networks and
\emph{AS} stands for the Internet Autonomous Systems with number
meaning the year data were gathered, \emph{Gorz\'ow Wlkp.},
\emph{{\L}\'od\'z} and \emph{Zielona G\'ora} are public transport
networks in corresponding Polish cities. $N$ is the number of
nodes, $\langle k \rangle$ - mean degree value, $c$ - clustering
coefficient. $A_e$ and $B_e$ mean experimental values (Fig.
\ref{fig_all}) whereas A and B are given by (\ref{e5}), A' and B'
by (\ref{e6}). $\Delta A$, $\Delta A'$, $\Delta B$, $\Delta B'$
stand for following differences $A-A_e$, $A'-A_e$, $B-B_e$,
$B'-B_e$.} \label{tab1}
\end{table*}

The influence of loops of the length three on the relation
(\ref{law0}) can be estimated as follows. Let us assume that in
the branching process forming the tree $T_i$ two nodes from the
nearest neighborhood of the node $i$ are {\it directly} linked
(the dashed line at Fig.\ref{fig_1}). In fact such a situation can
occur at any point of the branching tree $T_i$. Since such links
are useless for further network exploration by the tree $T_i$ thus
an {\it effective} contribution from both connected nodes to the
mean branching factor of the tree $T_i$ is decreased. Assuming
that clustering coefficients of every node are the same, the
corrected factor for the branching process equals to
$\kappa_c=\kappa-c\kappa$ where $c$ is the network clustering
coefficient. This equation is not valid for the branching process
around the node $i$ where $\kappa_i'=\kappa-c(k_i-1)$. A similar
situation arises around the node $j$. Replacing $k_i$ and $k_j$ with
$\langle k\rangle$ in $ \kappa_i'$ and $\kappa_j'$ one gets
\begin{equation}
k_ik_j[\kappa(1-c')]^{2}[\kappa(1-c)]^{x-3}=N\langle
k\rangle\label{prod2},
\end{equation}
where $c'=c(\langle k\rangle-1)/\kappa$. It follows that instead
of (\ref{e5}) we have
\begin{equation}\label{e6}
A'=3+\frac{\log(N\langle k\rangle) - 2\log[\kappa(1-c')]}{\log
[\kappa (1-c)]},\;\;\; B'=\frac{1}{\log[\kappa(1-c)]}\label{AB2}.
\end{equation}

The Table \ref{tab1} contains a comparison between the
experimental data (Fig. \ref{fig_all}) and theoretical predictions
given by Eq. (\ref{e5}) and (\ref{e6}). One can observe that the
range of parameters $A$ and $B$ for different networks is very
broad. Our approximate approach fits very well to random
Erd\H{o}s--R\'enyi graphs and BA models and is fairly good for
co-authorship and biological networks as well as for the Internet
Autonomous System and public transport network in Zielona G\'ora
while for two other transport systems it leads to larger errors.
Corrections due to clustering effects give a better fit for the coefficient $A'$,
while for some networks the coefficient $B$ is
closer to experimental value $B_e$ than $B'$.
The good agreement between theory based on random networks and empirical data
suggests that the considered real networks exhibit
a large level of randomness.

In conclusions we have observed universal path length scaling for
different classes of real networks and models. The mean distance
between nodes of degrees $k_i$ and $k_j$ is a linear function of
$\log (k_i k_j)$. The scaling holds over many decades regardless
of network degree distributions, correlations and clustering. A
simple model of random tree exploring the network explains such a
behavior and leads to a good agreement with experimental data. We
expect that the observed scaling law is universal for many complex
networks, with applicability reaching far beyond the quoted
examples.

\begin{acknowledgments}
We are thankful to S.N. Dorogovtsev for useful comments and to P.
W\'ojcicki for help with data collection. The work has been
supported by the KBN grant No 1P03B04727 and by the COST Action
P10 {\it Physics of Risk}. JAH is thankful to the Complex Systems
Network of Excellence {\it EXYSTENCE} for the financial support
during the Thematic Insitute at MPI-PKS Dresden.
\end{acknowledgments}



\begin{thebibliography}{7}
\bibitem{infoTran} In those networks vertices are bus- and tramstops while an edge exists if at least one public transport line crosses two stops. \cite{Julek}.
\end{thebibliography}
\end{document}